\documentclass[
 unsortedaddress,
 amsmath,amssymb,
 aps,
floatfix,
pra,
reprint,
prb,
]{revtex4-2}
\usepackage{caption}
\pdfoutput=1
\usepackage{subcaption}
\usepackage{graphicx}
\usepackage{dcolumn}
\usepackage{bm}
\usepackage{float}
\usepackage{romannum}

\usepackage{cleveref}
\usepackage{tablefootnote}
\usepackage{xcolor}

\begin{document}

\preprint{APS/123-QED}

\title{Understanding the Influence of Hydrogen on BCC Iron Grain Boundaries using the Kinetic Activation Relaxation technique (k-ART)}

\author{Aynour Khosravi}
 \email{aynour.khosravi@umontreal.ca} 
\affiliation{Département de Physique and Regroupement québécois sur les matériaux de pointe,\\ Université de Montréal, Montréal, Canada.}

\author{Jun Song}
 \email{jun.song2@mcgill.ca}
\affiliation{Department of Mining and Materials Engineering, McGill University, Montréal, Canada.}

\author{Normand Mousseau}
 \email{normand.mousseau@umontreal.ca}
\affiliation{Département de Physique and Regroupement québécois sur les matériaux de pointe,\\ Université de Montréal, Montréal, Canada.}

\date{\today}

\begin{abstract}

Hydrogen embrittlement (HE) poses a significant challenge to the mechanical integrity of iron and its alloys. This study explores the influence of hydrogen atoms on two distinct grain boundaries (GBs), $\Sigma37$ and $\Sigma3$, in body-centered-cubic (BCC) iron.  Using the kinetic activation relaxation technique (k-ART), an off-lattice kinetic Monte Carlo approach with an EAM-based potential, extensive catalogs of activated events for atoms in both H-free and H-saturated GBs were generated.
Studying the diffusion of H, we find that, for these systems, while GB is energetically favorable for H, this element diffuses more slowly at the GBs than in the bulk. The results further indicate that the $\Sigma 3$ GB exhibits higher stability in its pure form compared to the $\Sigma 37$ GB, with notable differences in energy barriers and diffusion behaviors. Moreover, with detailed information about the evolution landscape around the GB, we find that the saturation of a GB with hydrogen both stabilizes the GB by shifting barriers associated with Fe diffusion to higher energies and reducing the number of diffusion events. For the $\Sigma 37$ GB, the presence of hydrogen causes elastic deformation, affecting the diffusion of Fe atoms both at the GB and in adjacent positions. This results in new diffusion pathways but with higher diffusion barriers, unlike for the $\Sigma 3$ GB. These results indicate that the presence of hydrogen rigidifies the direct GB interface layers while allowing more atoms to be active for the $\Sigma 37$ GB. This provides a microscopic basis to support the existence of competing mechanisms compatible with either plasticity (such as hydrogen enhanced localized plasticity --- HELP) or energy-dominated (hydrogen enhanced decohesion mechanism ---HEDE) embrittlement, with the relative importance of these mechanisms determined by the local geometry of the GBs.

\end{abstract}

\maketitle
\pagenumbering{arabic}

\section{Introduction}

Hydrogen (H) has long been recognized as a factor that leads to the deterioration of the mechanical capabilities of metals, presenting a significant technological hurdle in various industrial applications~\cite{gangloff2012gaseous}. Specifically, the diffusion and segregation of hydrogen in iron and its alloys contribute to engineering challenges related to embrittlement (HE), and deterioration of high-strength steels, and similar components. Hydrogen, a prevalent impurity in iron-based materials, is integrated into materials during both production and service, exhibiting notable mobility within the bulk phase. Research further indicates that microstructural defects in the material, such as vacancies, dislocations, and grain boundaries (GB), can effectively trap hydrogen impurities~\cite{novak2010statistical, ramasubramaniam2008effect, wen2003atomistic, zhong2000charge, abdalla2018hydrogen, sun2021current}. 

Despite the ambiguity surrounding the precise origin of hydrogen embrittlement and the diverse factors that ultimately influence this behavior, four main HE mechanisms have been proposed: (i) hydrogen enhanced localized plasticity (HELP); (ii) hydrogen enhanced decohesion mechanism (HEDE); (iii)hydrogen enhanced strain-induced vacancy formation (HESIV); and  (iv) adsorption-induced dislocation emission (AIDE)~\cite{nagumo2016fundamentals}. Understanding which HE mechanism contributes to material failure, or whether a synergy of mechanisms is at play, remains an open question for which numerical methods could provide additional information and clarity. 

The relative importance of these HE mechanisms is believed to depend on several key factors: hydrogen concentration, crystal structure (BCC, FCC, or HCP) and microstructure of the alloys. In general, HE mechanisms can be divided into two categories: ``plasticity-mediated'' and ``hydrogen-impeded plasticity''~\cite{djukic2019synergistic}. ``Plasticity-mediated'' HE mechanisms, such as HESIV and HELP, are activated through the interaction of hydrogen with vacancies and dislocations. 
In contrast, the HEDE mechanism falls under ``hydrogen-impeded plasticity'' and predominantly occurs at GBs, twin boundaries (TBs), phase interfaces, and matrix/precipitate interfaces. 

To better understand hydrogen-induced intergranular cracking, considerable numerical effort has been dedicated to studying the trapping and diffusion behaviors of hydrogen atoms at microstructural defects, particularly at grain boundaries. This knowledge is critical for developing materials that are resistant to hydrogen embrittlement. Numerous studies~\cite{bechtle2009grain, wan2019hydrogen, yamaguchi2012mobile, wu1994first} demonstrate that GBs play a significant role in HE. It is important to study the key role of GBs in hydrogen diffusion because they can either slow down or speed up hydrogen diffusion depending on the conditions~\cite{PhysRevLett.122.215501}. This behavior can be explained by two opposing theories: short-circuit diffusion and boundary-trapping diffusion. Short-circuit diffusion theory suggests that hydrogen atoms can move faster along grain boundaries than through the lattice, thus accelerating hydrogen transport in metals.
By comparing two kinds of grain boundaries in FCC and BCC structures, Lee~\textit{et al.}~\cite{LEE1986301} and Mine~\textit{et al.}~\cite{MINE20118100} concluded that grain boundaries in BCC structures can act as trapping sites for hydrogen atoms, while grain boundaries in FCC structures can serve as fast migration channels. This difference is due to the higher density of grain boundaries in BCC structures, which increases the likelihood of hydrogen trapping at dislocations. However, a DFT (Density functional theory) study by Du~\textit{et al.}~\cite{du2011first} examining hydrogen interactions with GB structures in $\alpha$-and $\gamma$-Fe found that none of the examined GBs exhibited fast hydrogen diffusion and the hydrogen diffusivity within these GBs was lower compared to diffusion in perfect single-crystalline bulk, causing these GBs to act as hydrogen traps and lowering the critical strain required for material fracture. The accommodation of hydrogen depends on the local coordination of interstitial sites, with larger interstitial sites in the open-grain boundary structures enhancing solubility. This highlights the fact that there is no consensus on this topic and that further studies are needed to determine the exact impact of GB on hydrogen atoms.

To study the diffusion of GBs in the presence of H, it is crucial to consider the effect of H atoms on the diffusion of atoms within GBs. However, much less atomistic simulation work, has focused on how the presence of hydrogen can alter the diffusivity of the GB itself by changing the potential energy surface, which is related to deformations around the GB. It is understood that the strength, deformability, and fracture toughness of structural materials such as iron and iron alloys are likely influenced by GBs. Although experimental findings suggest that hydrogen can affect dislocation plasticity and the transition of fracture modes, delving into nanoscale interaction mechanisms between hydrogen and GBs has been more appropriately addressed through theoretical simulations~\cite{ding2021hydrogen}. 
The observations of intergranular fracture surfaces suggest that hydrogen weakens GB strength, with higher hydrogen concentrations at the GBs correlating with lower GB cohesion. 
The activation of the HEDE mechanism and the initiation/propagation of hydrogen-assisted cracks are controlled by reaching a critical local hydrogen concentration, which weakens the interatomic forces. However, the precise quantitative relationship between the local hydrogen concentration and the consequent reduction in GB strength remains experimentally elusive~\cite{PhysRevMaterials.1.033603, wasim2021influence, djukic2019synergistic}.
Therefore, it is crucial to examine the interactions between hydrogen and GBs, particularly under conditions with a sufficient concentration of hydrogen atoms. This understanding is essential for elucidating the exact mechanism of HEDE, which significantly affects the mechanical properties of the materials. While the topic of hydrogen-GB interactions have been extensively studied using Density Functional Theory (DFT) and Molecular Dynamics (MD) simulations ~\cite{rice1989embrittlement, kirchheim2015chemomechanical, yamaguchi2012mobile, he2021first, du2011first, yamaguchi2011first, jung2018influence, liu2011effects, wang2016effect, song2013atomic}, there has been a lack of focus on scenarios involving the impact of H atoms on atoms within GBs when there is an adequate hydrogen concentration in the GB. This gap underscores the need for more detailed investigations into hydrogen segregation in GBs and the interaction of hydrogen atoms with these boundaries to achieve a deeper understanding of these mechanisms.

In this work, we focus on the more specific question of the effect of hydrogen on the kinetics of the Fe atoms at the GB, an aspect that has received relatively little attention from computer simulations over the year. To do so, we perform atomistic simulations on two GBs using empirical potentials with the kinetic Activation-Relaxation Technique (k-ART), an off-lattice kinetic Monte-Carlo algorithm with on-the-fly construction of the event catalog.  This technique allows us to follow long-time kinetics as well as obtain detailed mapping of the energy landscape around GBs. These results help to understand the intricate dynamics of hydrogen incorporation and its influence on the behavior of Fe at the GBs.

The organization of this paper is as follows. We begin with an overview of the methodologies employed in this study, as well as detailed simulation details (Section~\ref{section:Methodology}). Section~\ref{section:Results} presents the migration energies of Fe atoms across two different GB types, together with an examination of the impact of hydrogen atoms on these boundaries. Finally, we discuss the results and their significance in the context of hydrogen-assisted defect mobility.

\section{Methodology}
\label{section:Methodology}
\subsubsection{Models of grain boundaries}

Our study focuses on two separate GBs each in a box with periodic boundary conditions using crystalline BCC Fe and relaxed at zero pressure at $T=0$~K. These configurations feature two symmetric tilt GBs, with $\langle 100 \rangle$ or $\langle 110 \rangle$ axes of rotation parallel to the GB plane, constituting high-angle boundaries with misorientations exceeding $15^{\circ}$. The first sample is of the $\langle 100 \rangle$ family, type $\Sigma 37(160) \theta = 18.93^{\circ}$, and the second GB belongs to the $\langle 110 \rangle$ family, type $\Sigma 3(112)\theta=70.53^{\circ}$.

To maintain consistency with the existing literature, cells containing GBs are selected from Ref.~\cite{PhysRevB.89.014111}.
For the $18.93^{\circ}$ $\langle 100 \rangle$  model, the cell box dimensions are $17.140 \times 51.878 \times 51.686$~\AA$^{3}$ with 3924 atoms; for the $70.53^{\circ}$ $\langle 110 \rangle$ model, the box measures $36.318 \times 34.557 \times 37.367$~\AA$^{3}$ with 4032 atoms.
To ensure adherence to the periodic boundary conditions in all directions, the two GBs are positioned at a distance equivalent to half the box size along the z-direction.

The $18.93^{\circ}$ $\langle 100 \rangle$ structure represents a general GB with medium GB energies $(\gamma_{GB} = 61.05  ~meV/$\AA$^{2})$, chosen for its distinct structural characteristics and potential to trap H atoms; the $70.53^{\circ}$ $\langle 110 \rangle$ GB with $\gamma_{GB} = 16.22  ~meV/$\AA$^{2}$, extensively studied experimentally~\cite{lejcek2010grain}, features a unique configuration belonging to $\Sigma 3$GBs, because of its singular structural unit, it results in a pronounced decrease in GB energy. 

\subsubsection{Computing the energy of a grain boundary and Interatomic potential}

The energy of a GB ($\gamma_{GB}$) is determined by subtracting the total energy of a supercell with the GB $(E^{GB}_{tot})$ from the reference energy of the same number of atoms in the bulk $(E^{bulk}_{tot})$. This reference energy is calculated using a bulk unit cell or a supercell oriented similar to the GB, which represents the energy change caused by the presence of the GB in the system. To find the energy per unit area of the GB, the energy difference is divided by twice the area of the GB plane $(2A)$, because there are two GBs in the simulation box to have a periodic structure in all three directions:
\begin{equation}\label{eq:3}
    \gamma_{GB} = \frac{E^{GB}_{tot} - E^{bulk}_{tot}}{2A}.
\end{equation}
Note that this quantity plays an important, but not unique, role in defining the stability and behavior of GBs.

The solution energy of an interstitial solute atom in the GB or bulk (calculated using Eq.~\ref{eq:4}) refers to the energy required to incorporate a solute atom into a specific structure. The solution energy in a GB is represented by $E^{GB}_{sol}$ and, in the bulk, by $E^{Bulk}_{sol}$. It represents the difference in energy between the GB or bulk with the interstitial solute atom and the pristine GB (or bulk) without it. This energy arises because of the distortion of the GB structure and the interactions between the solute atom and the surrounding atoms in the GB:
\begin{equation}\label{eq:4}
    E^{GB/Bulk}_{sol} = \frac{E^{+nH}_{tot} -( E^{pure}_{tot} +N_{x}) \mu_{H}}{n_H}.
\end{equation}

In this equation, $E_{tot}^{+nH}$ denotes the overall energy of either the bulk or the system with GB, augmented by `$n$' instances of the solute atom H. $E_{tot}^{pure}$ represents the complete energy of the bulk or of the pristine GB structure, where only the matrix atoms contribute. Finally, `$\mu_{H}$' denotes the reference chemical potential, computed for the solute atom H.

The segregation energy can be derived by calculating the solution energies of both the bulk and GB structures. Negative segregation energies indicate that impurity atoms tend to segregate towards the GB plane,
\begin{equation}\label{eq:5}
    \gamma_{seg}^{GB} = E^{GB}_{sol} - E^{Bulk}_{sol}.
\end{equation}

In the case of interatomic potentials, Fe-H models incorporating FeH and H-H components are generally based on Mendelev's adaptation of the Embedded Atom Method (EAM) potential for Fe~\cite{mendelev2003development,ackland2004development}. Due to their focus on elemental hydrogen, some of these H potentials can lead to the nonphysical clustering of interstitial H atoms~\cite{ramasubramaniam2010erratum}. To avoid these issues, we select a Finnis-Sinclair type EAM potential for this study, with Fe-H parameters taken from Song \textit{et al.}~\cite{song2013atomic} and Ramasubramanian \textit{et al.}~\cite{ramasubramaniam2010erratum}. These parameters were adjusted to accurately simulate the Fe-H interactions in bulk Fe while minimizing the unphysical aggregation of H atoms. The forces and energies are calculated from LAMMPS' implementation of these potentials~\cite{plimpton1995fast, sandiaLAMMPSMolecular} and are linked to k-ART by treating LAMMPS's as a library. Our previous study~\cite{khosravi2023kinetics} found that while this potential introduces a shallow metastable state along the defect diffusion pathway~\cite{starikov2021angular}, other defect related properties~\cite{song2013atomic} are well reproduced, as confirmed by ab initio calculations~\cite{khosravi2023kinetics}.

\subsubsection{The kinetic activation-relaxation technique (k-ART)}

Computational approaches and atomistic simulations play a crucial role in understanding the microscopic processes linked to atomic diffusion, with advancements in methodologies and increased computing capabilities.

The simulations in this study are carried out using the kinetic Activation-Relaxation Technique (k-ART)~\cite{el2008kinetic, beland2011kinetic}, an off-lattice kinetic Monte Carlo method (KMC) that generates activated events around certain configurations using the activation-relaxation technique nouveau (ARTn) method~\cite{barkema1996event, malek2000dynamics, mousseau2012activation}, and NAUTY, a topological analysis package, as the generic classification method~\cite{mckay1981practical}. We provide a brief overview of the fundamental algorithm employed in the k-ART method, along with its associated parameters. Using NAUTY, a topological analysis tool developed by McKay~\cite{mckay1981practical}, within a locally relaxed system, we determine the local topology surrounding each atom. This involves generating a graph encompassing all atoms within, for these systems, a 6~Å radius of the central atom, corresponding to a cutoff between the 6th and 7th neighboring shell and including approximately 65 atoms, with vertices connecting atoms within 2.7~Å of each other, a distance between the first and the second neighbor shell. Subsequently, this constructed connectivity graph undergoes analysis by NAUTY, which provides a unique identifier that characterizes its automorphic group, inclusive of its chemical identity. Atoms that share the same topological environment exhibit similar activated mechanisms~\cite{beland2011kinetic}. This hypothesis is examined and corrected by adjusting in various thresholds as needed.
For each topology, an adequate number of ARTn searches are performed to detect associated events~\cite{barkema1996event, malek2000dynamics, machado2011optimized, beland2011kinetic}. This process involves deforming the local environment surrounding a selected atom in an arbitrary direction until the system partially relaxes, as indicated by the lowest eigenvalue of the Hessian matrix becoming negative. Following this, the configuration is pushed along the direction of negative curvature until a first-order saddle point is reached, which is characterized by the total force falling below a predefined threshold value near zero. Subsequently, the system is pushed over the saddle point and relaxed into a new minimum. All events are checked for connectivity, meaning that before adding the event to the catalog, we ensure that the saddle point is also connected to the initial minimum.

In this study, each new topology starts with 50 independent ARTn searches. To ensure that frequently found environments are well sampled, the number of searches launched increased proportionally to the logarithm of the number of times a topology is encountered, as follows :
\begin{equation}\label{logsearch}
   N_{\mathrm{search}} = \max[S_F  (1 + \log_{10} C)- A,0]
\end{equation}
\noindent
where $N_{\mathrm{search}}$ defines the number of new searches, $S_F$ is the basic number of searches (here, 50), $C$ represents the number of times the topology has been observed, and $A$ refers to the previous number of attempted event searches initiated. This logarithmic search function ensures that the topologies encountered more frequently are subjected to additional search.
When an event is added to the database, it is also included in the binary tree of the events and histograms. After the catalog is fully updated and the tree is constructed for the current atomistic configuration, generic events are ranked according to their rate:
\begin{equation} \label{eq:1}
\Gamma(i) = \nu_{0} exp \left( {-\frac{E_b(i)}{k_B T}} \right),
\end{equation}
where $E_b(i)$ represents the activation (barrier) energy for event $i$, which is measured as the difference between the transition state and the initial minimum, and $\nu_0$ denotes the attempt frequency (prefactor). While k-ART can compute this prefactor for each event using the harmonic approximation, we found previously that the prefactor does not vary much for H in Fe~\cite{khosravi2023kinetics} and we fix it at $10^{13} Hz$ in the simulations presented here. 

Events are sorted based on the initial rate assessment, and those exceeding a minimum probability threshold (in this case, 1 in 10,000 or higher) undergo complete specific reconstruction and reconvergence to include local and nonlocal elastic deformations. Following this process, the total rate is reassessed.

KMC time steps are determined based on a Poisson distribution,
\begin{equation}\label{eq:2}
    t = - \frac{\ln \mu}{\sum_i \Gamma_i},
\end{equation}
where $\mu$ is a uniformly random number distributed between $[0,1]$ and $\Gamma_i$, which is the rate of each event attainable within the configuration.

In this study, all simulations are performed at a constant temperature of 300~K for kinetic Monte Carlo evaluation. The energy landscape remains unaffected by temperature variations when barriers are significantly higher than the temperature; therefore the temperature selection only impacts the selected barriers and not the event catalog.

\section{Results}
\label{section:Results}
We first establish the behavior of the two symmetric $\Sigma 37$ and $\Sigma 3$ grain boundaries in the absence of H. Initially, we examine the behavior of the GB in the absence of hydrogen. Subsequently, we analyze the kinetics of a single hydrogen atom within the GB. Finally, we investigate the kinetics of Fe atoms surrounding the GB after saturation with H atoms.

\subsubsection{$\Sigma 37$ GB }

\begin{figure}[!tb]
    \centering
    \includegraphics[scale=0.52,angle=270]{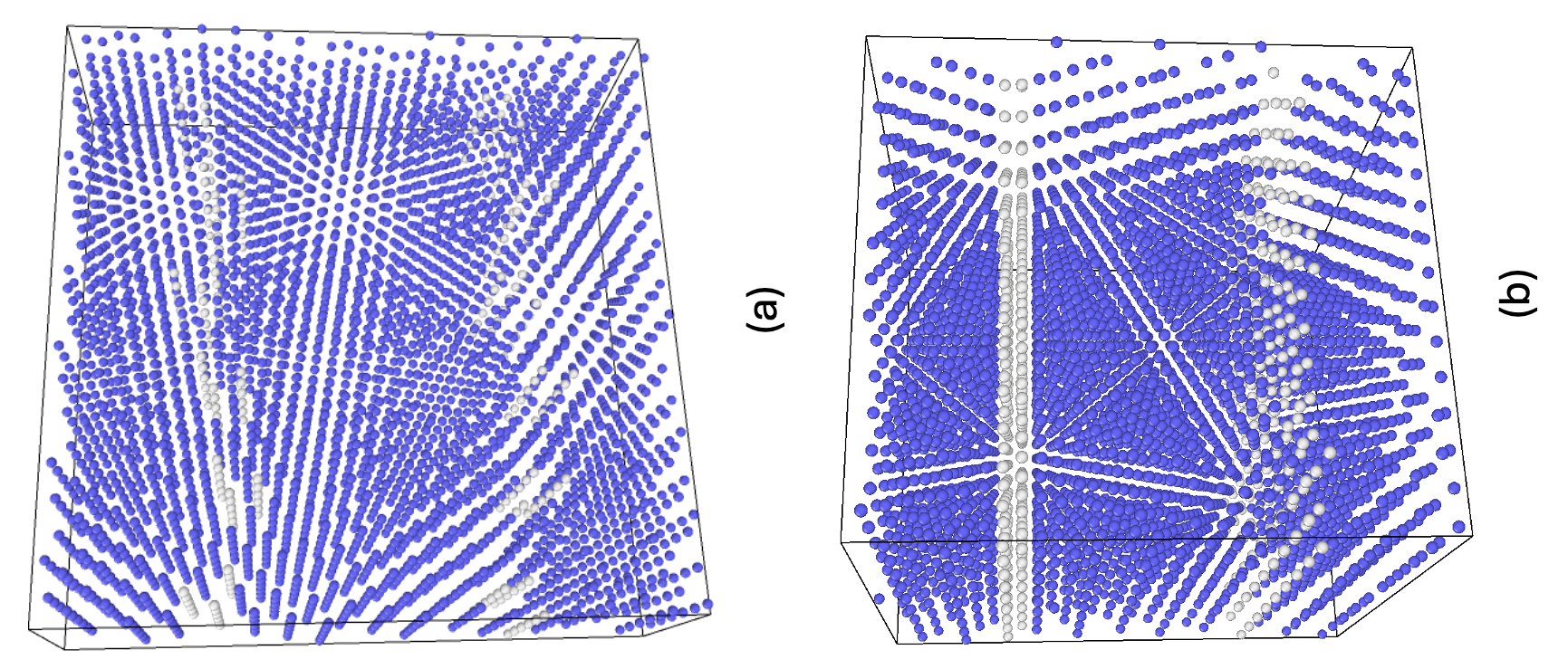}
    \caption{Representation of the pure  (a) $\Sigma 37$ and (b) $\Sigma 3$ GB structures. Blue atoms denote iron (Fe) atoms arranged in a body-centered cubic (BCC) structure, while white atoms represent Fe atoms associated with the GB without a BCC structure.}
    \label{fig1}
\end{figure}

The $\Sigma 37(160)$ symmetric tilt GB studied here is characterized by a tilt angle of $18.93^{\circ}$. Fig.~\ref{fig1} (a) presents an atomistic depiction of the GB that represents its structural features. 


The structural stability of the $\Sigma 37$ GB in the absence of hydrogen was previously explored by Restrepo \textit{et al.}~\cite{PhysRevB.97.054309} using k-ART. This work confirmed that the $18.93^{\circ}$$\langle 100 \rangle$ GB system is very stable at room temperature, with a dynamics dominated by transitions between the ground state and less stable higher energy configurations situated 1.4 to 1.6~eV above the ground state. Notably, these higher-energy states were found to be inherently unstable and necessitated overcoming a minor energy barrier to revert to the ground-state arrangement.

The application of k-ART allows the generation of a large set of activated mechanisms that are necessary to characterize the behavior of Fe atoms surrounding the GB through a detailed classification of their dynamics. Using k-ART, we generate an extensive event catalog that offers insights into the intricate interplay between Fe atoms in the GB environment, helping to understand how the presence of H alters the kinetics of Fe atoms within this context. The purpose of this study is to document the various diffusion mechanisms of Fe atoms within the GB region, providing a solid foundation for further analysis of the kinetics of Fe and H.  

\begin{figure}[!tb]
    \centering\includegraphics[scale=0.26]{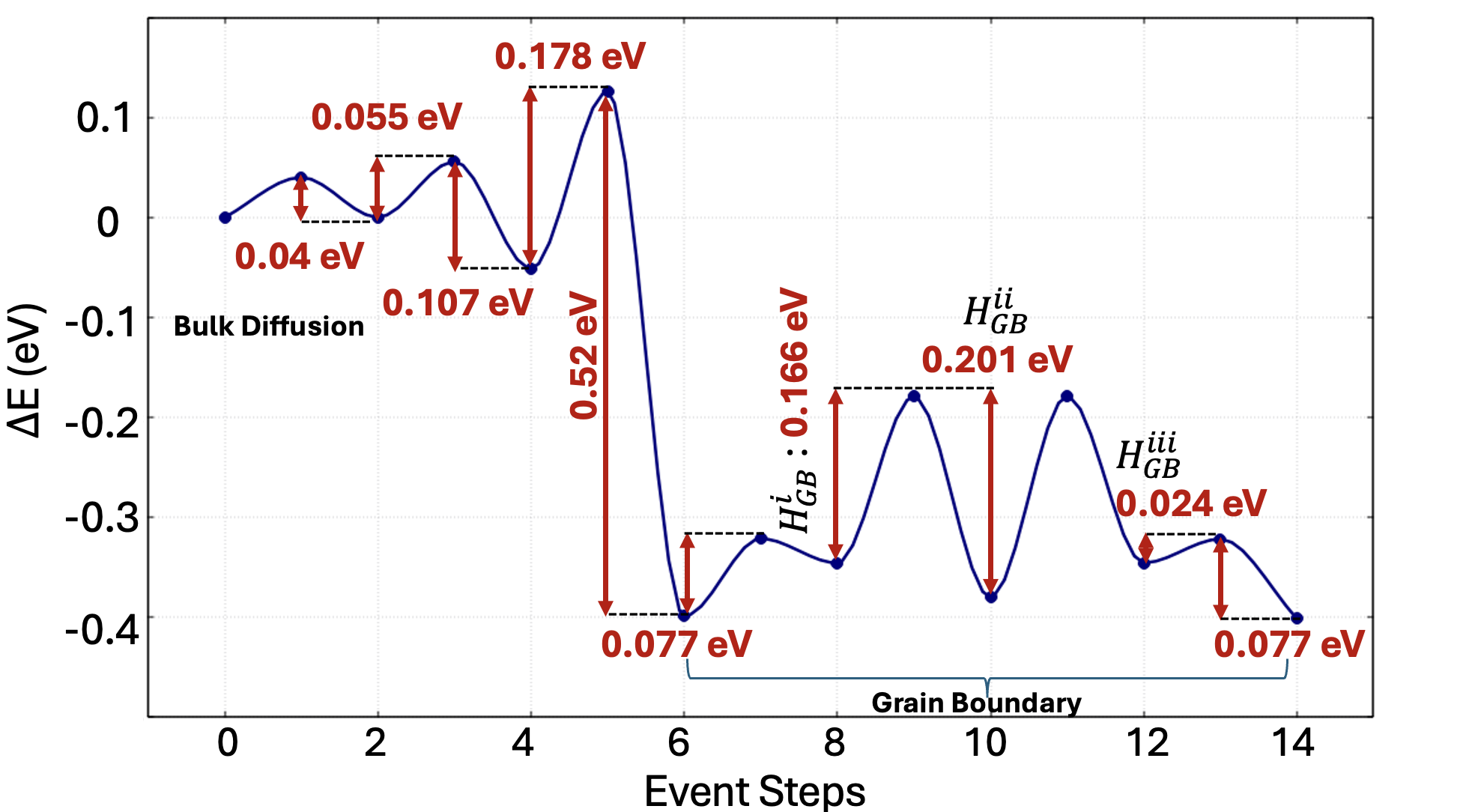}
    \caption{Illustration of the potential energy surface of a single hydrogen atom within and surrounding the $\Sigma 37$ GB. Three successive steps are required for hydrogen trapping within the GB from the bulk.}
    \label{Trapping1H_sigma37}
\end{figure}

First, we inset a single H atom within the$\Sigma 37$ GB. GBs represent particularly favorable environments for H accumulation because of their non-crystalline nature. We initiate the simulation by placing a single H atom in the bulk, at the center of the simulation box, where there is neither elastic deformation nor GB influence. This allows us to observe the diffusion of H within the box and its eventual trapping within the GB. The pathway for H atom diffusion and trapping in GB is described in Fig.~\ref{Trapping1H_sigma37}. The initial phase, shown in Fig.~\ref{Trapping1H_sigma37}, focuses on the last steps of H diffusion from the bulk into the GB. As in Ref.~\onlinecite{khosravi2023kinetics}, H diffusion in the bulk is controlled by a 0.04~eV barrier. The effects of the GB on H are short-range and become noticeable only in the third-neighbor position. For H to become fully trapped within the GB, it must overcome barriers of 0.055~eV, 0.178~eV, and 0.077~eV successively. Although a lower barrier for bulk diffusion may seem preferable for insertion into the GB, these barriers are still accessible even at room temperature, as indicated by their respective event rates: 1.153$\times 10^{+12}$ Hz for 0.055~eV, 1.031$\times 10^{+10}$ Hz for 0.178~eV, and 4.992 $\times 10^{+12}$ Hz for 0.077~eV.

As a result, H diffusion mechanisms within the $\Sigma 37$ GB can be effectively classified into two distinct groups: those associated with tunnel diffusion within the GB ($H^{\romannum{1}}_{GB}$, $H^{\romannum{2}}_{GB}$ and $H^{\romannum{3}}_{GB}$) and those linked to detrapping from the GB ($H^{\romannum{1}}_{Detrapp}$, $H^{\romannum{2}}_{Detrapp}$ and $H^{\romannum{3}}_{Detrapp}$). As shown in Table~\ref{tab:table2}, the barriers observed for the $H^{\romannum{1}}_{GB}$ - $H^{\romannum{2}}_{GB}$ - $H^{\romannum{3}}_{GB}$ mechanisms exhibit smaller magnitudes at 0.166, 0.201, and 0.024~eV, for an effective diffusion barrier of 0.22~eV the GB. Detrapping occurs through a multiple-step process characterized by higher effective barriers, at 0.384, 0.442, and 0.655~eV; therefore, the rates are lower. Although all of these barriers are accessible at room temperature, the smaller ones are notably more likely to occur, indicating that, statistically, H diffuses within the GB before it hops back into the bulk. Note that hydrogen moves significantly slower within this GB than in the bulk. 

\begin{table}[!tb]
\caption{\label{tab:table2}
List of the barriers for the diffusion of a  $H$ atom in the the $\Sigma 37$ GB structure (see text for the definition of the various events).
The barrier is calculated as the energy difference between the saddle point and the initial state ($E_b = E_{sad} - E_{init}$). The inverse barrier represents the energy difference between the final state and the saddle point($E_{inv} = E_{final} - E_{sad}$). Rates are computed using \Cref{eq:1}. Root-mean square displacement $d_{si}$ ($d_{fi}$) between the saddle point (final minimum) and the initial state are also indicated.
}
\begin{ruledtabular}
\begin{tabular}{cccccc}
\scriptsize{ \begin{tabular}{@{}c@{}} Diffusion \\ Mechanism \end{tabular}} &\scriptsize{ \begin{tabular}{@{}c@{}}Barrier \\ (eV)\end{tabular}}& \scriptsize{ \begin{tabular}{@{}c@{}}Inverse \\ Barrier (eV)\end{tabular}}& $\Gamma_{if}$ (Hz) & $d_{si}$ 
(\AA) & $d_{fi}$ (\AA)\\
\hline
\\
$H^{\romannum{1}}_{GB}$ & 0.166 & 0.201 & 1.593$\times 10^{+10}$ & 0.764  &  1.657 \\
\\
$H^{\romannum{2}}_{GB}$ & 0.201 & 0.166 & 4.191$\times 10^{+09}$  & 0.930  &  1.657 \\
\\
$H^{\romannum{3}}_{GB}$ & 0.024 & 0.077 & 3.925$\times 10^{+12}$& 0.514 &  1.388 
\\
\\
\hline
\\
$H^{\romannum{1}}_{Detrapp}$ & 0.384 & 0.052 & 3.476$\times 10^{+06}$  & 1.312  &  1.73 \\
\\
$H^{\romannum{2}}_{Detrapp}$ & 0.442 & 0.052 & 3.650$\times 10^{+05}$  & 1.415  &  1.747 \\
\\
$H^{\romannum{3}}_{Detrapp}$ & 0.656 & 0.163 & 9.644$\times 10^{+01}$  & 1.365  &  1.99 \\
\\
\end{tabular}
\end{ruledtabular}
\end{table}

With the knowledge of the energy landscape of a single H, we now turn to the H-saturated GB. To establish the saturation concentration for H atoms within the GB, we perform NVT molecular dynamics simulations at room temperature for 30 ns using LAMMPS~\cite{sandiaLAMMPSMolecular}, relaxing the initial and final configurations with different H concentrations to zero pressure. The concentration of H was progressively increased until saturation within the GB is reached. Our results reveal that the optimal saturation level is achieved when hydrogen atoms make up 3.54 at.\%  (144 H atoms for 3924 Fe) of the total structure; as additional H atoms are introduced, H atoms show barrierless detrapping from the GB and diffuse almost immediately into the bulk.

The interface energy of the GB, the solution energy of hydrogen atoms throughout the bulk and GB, and the segregation energy of the GB are presented in Table~\ref{tab:table3}. The H segregation energy is negative for the $\Sigma 37$ GB, indicating a propensity for atom segregation towards the GB interface.

\begin{table}[!tb]
\caption{\label{tab:table3}
The GB interface energy, hydrogen atom solution energy in bulk and at GBs, and segregation energy of GB calculated for the two GBs studied here.
}
\begin{ruledtabular}
\begin{tabular}{ccccc}
 &  \scriptsize{ \begin{tabular}{@{}c@{}}Interface energy \\ $\gamma_{GB}$ \\ ({$eV$}/{$\AA^2$}) \end{tabular}} & \scriptsize{ \begin{tabular}{@{}c@{}}$E^{GB}_{sol}$ \\ ($eV$) \end{tabular}} &  \scriptsize{ \begin{tabular}{@{}c@{}}$E^{Bulk}_{sol}$\\ ($eV$) \end{tabular}} &  \scriptsize{ \begin{tabular}{@{}c@{}} Segregation energy \\ $\gamma^{GB}_{seg}$ \\ ($eV$) \end{tabular}} \\
\hline
\\
$\Sigma 37$ GB & 0.061  & 0.925 & 1.307 & -0.382 \\[1em] 
$\Sigma 3$ GB & 0.162 & 2.291 & 1.307 & 0.98 \\[1em] 
\end{tabular}
\end{ruledtabular}
\end{table}

To understand the impact of H-saturation on the stability of the Gb, we first establish a baseline by examining the kinetics of the Fe atoms around the GB in the pure system, and then proceed to the H-saturated system. 

\begin{figure*}[!tb]
    \centering
    \includegraphics[scale=0.33]{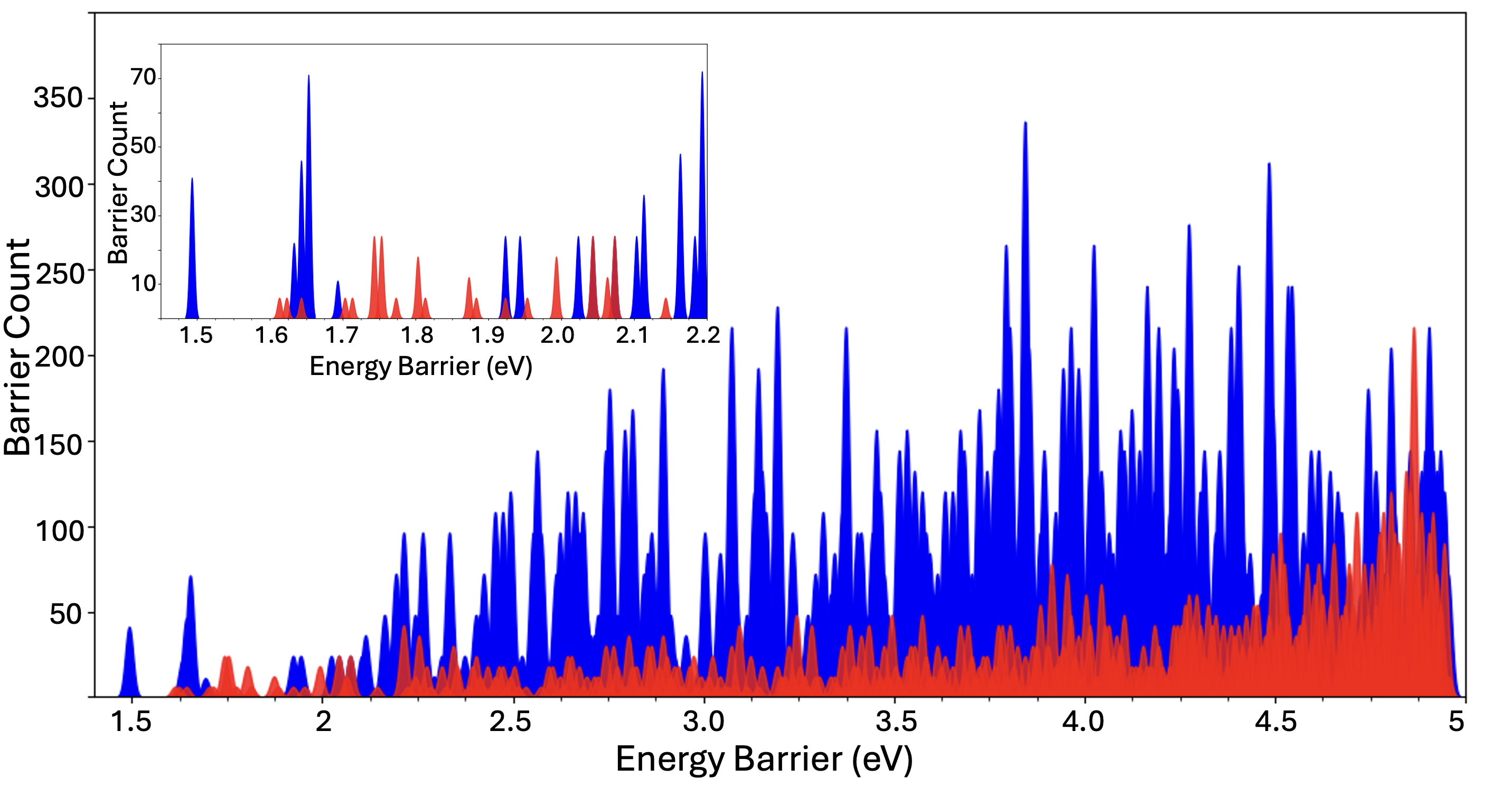}
    \caption{Energy-barrier distribution of the 21,482 events centered on Fe generated in the pure $\Sigma 37$ GB  and of the 8,368 events also centered on Fe but generated in the H-saturated GB as a function of the energy barrier. Diffusion barriers for Fe atoms in the pure system are represented in blue, while red represent barriers for the H-saturated GB. Inset: zoom on the lowest energy part of the distribution.}
    \label{sigma37_distribution}
\end{figure*}

Using k-ART, we identify the activated barriers for all 3924 Fe atoms within the $\Sigma 37$ GB, that is, all those that are not in a BCC environment. As discussed in the method, for every identified topology label, k-ART initiated 50 sets of ART nouveau searches to detect all activated occurrences, increasing the number of searches as certain topologies become more prevalent. In the absence of hydrogen, we observe 60 distinct topologies in the initial structure, including crystalline topologies, whereas 82 unique topologies were distinguished in the hydrogen-saturated system.

Fig.~\ref{sigma37_distribution} presents, the distribution of the Fe diffusion barriers for pure GB in blue. Launching searches from 60 initial topologies, generated 21,482 different events. In red, we show the distribution of the 8,368 events with Fe-dominated diffusion of the H-saturated GB (events where H moves with significant Fe diffusion are ignored here). 

\begin{figure}[!tb]
    \centering
    \includegraphics[scale = 0.37]{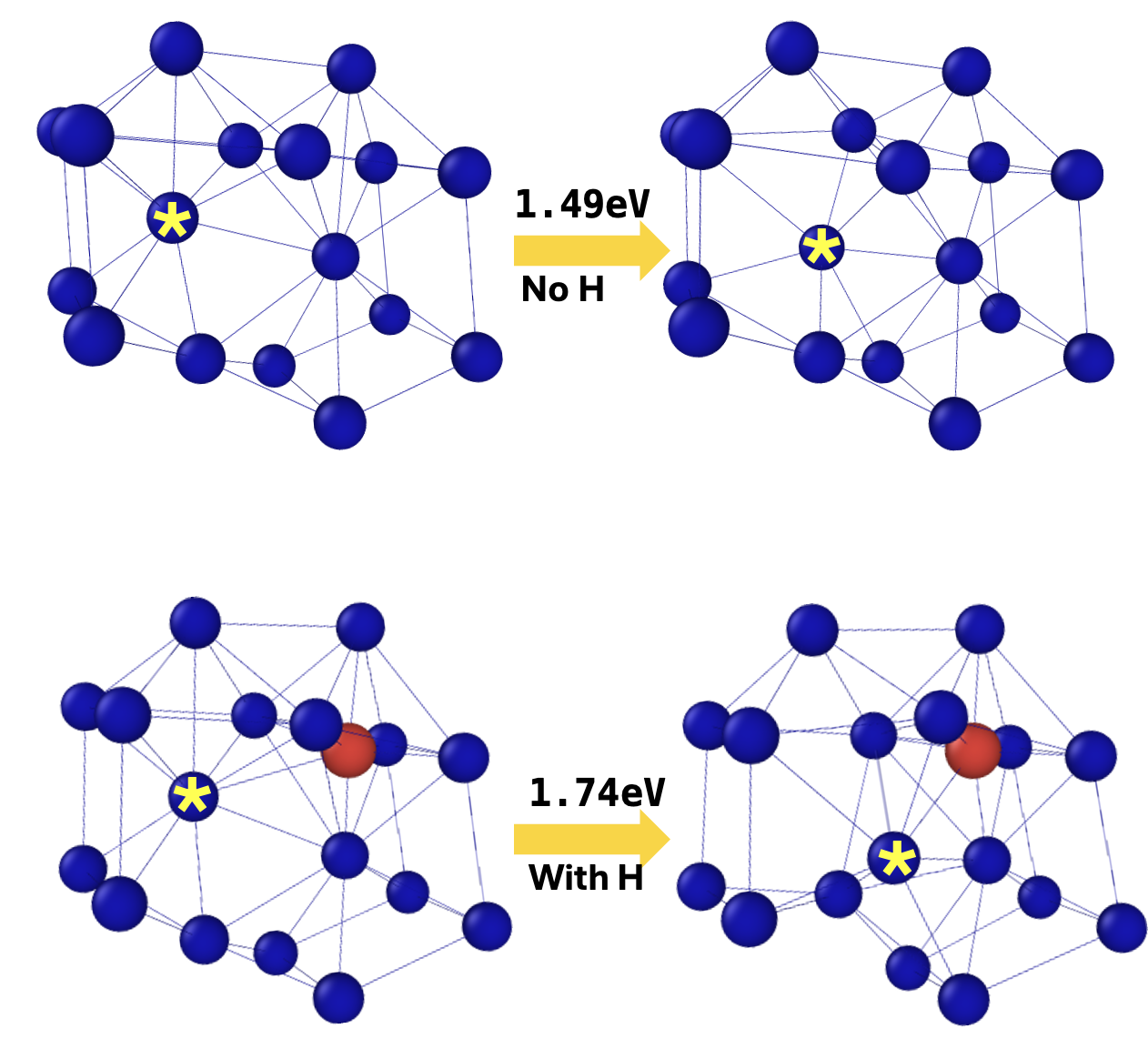}
    \caption{This figure illustrates the topology of the same atoms for the pure GB and the GB saturated with hydrogen (H). Iron atoms are depicted as blue spheres, while hydrogen atoms are represented by red spheres. The barrier associated with the displaced iron atom is denoted with an asterisk (*). The barrier is shifted when comparing these two scenarios.}
    \label{SameAtomEvent}
\end{figure}

Although the difference in the number of events found is difficult to link to specific physical effects, referring to the distribution shown in Fig.~\ref{sigma37_distribution}, we note a number of changes. First, the total number of events is significantly reduced. Second, the barrier distribution shifts to a higher energy when the GB is saturated and modified. Focusing on the low-energy spectrum, from 1.4 to 2.2~eV (see Inset), for example, the lowest-energy Fe-diffusion mechanisms in the pure system, at 1.49~eV and around 1.64~eV, are pushed to higher energy: that the lowest energy barrier, at 1.49~eV, disappears in the H-saturated system, while those between 1.63- and 1.68~eV appear are much less probable and are shifted in energy. A more detailed look at specific mechanisms shows that the first really diffusive barrier for Fe in the H-saturated GB is at 1.74~eV, a shift of 0.25~eV with respect to the pure system. The topologies associated with this event in both systems are illustrated in Fig.~\ref{SameAtomEvent}. In this figure, the atom whose barrier is affected is marked with an asterisk (*). This indicates that the saturation of H atoms in the system alters the diffusion barrier for Fe atoms.

\begin{figure}[!tb]
    \centering
    \includegraphics[scale = 0.36]{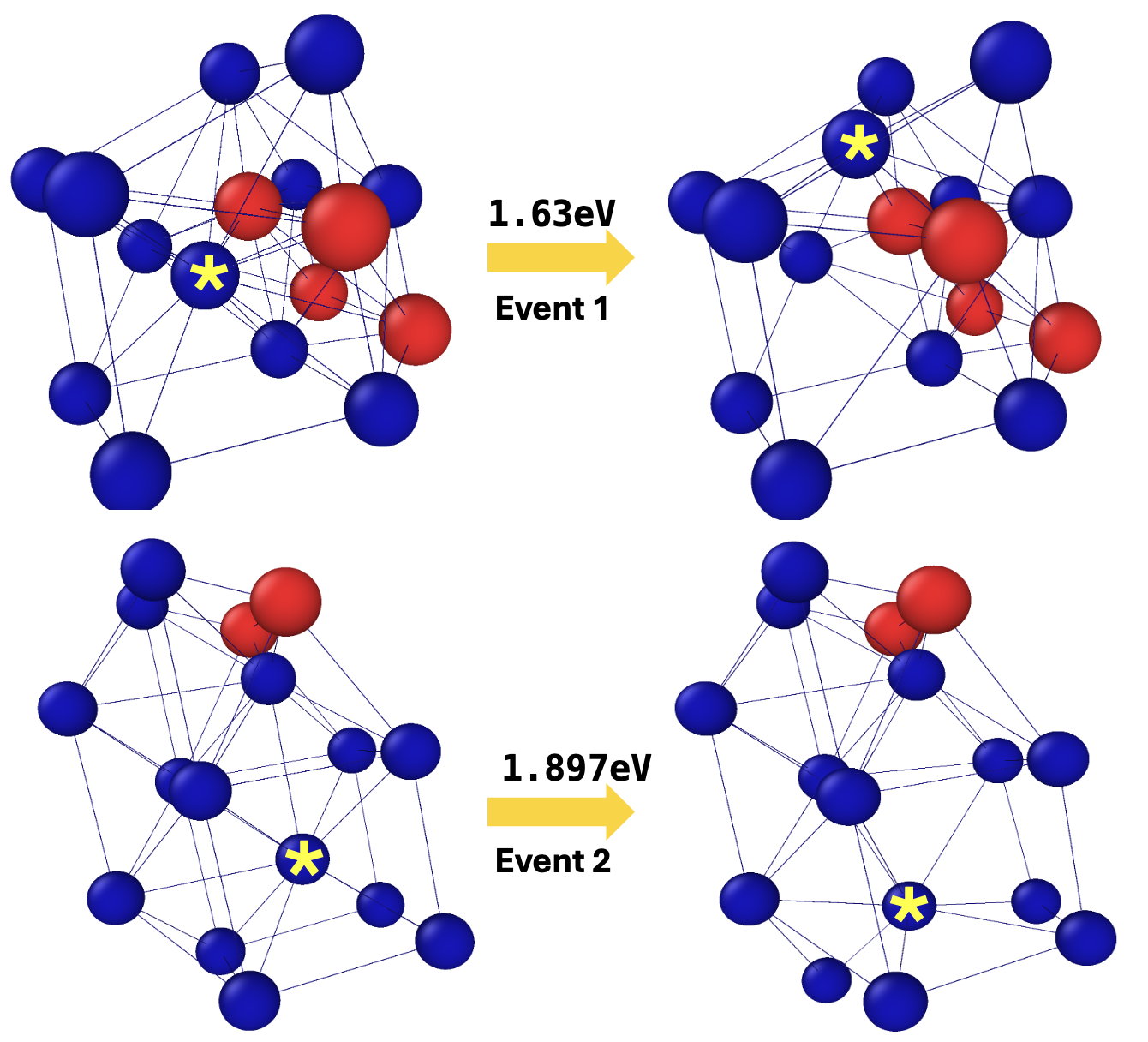}
    \caption{Two examples of the low energy barrier mechanisms(between 1.6 and 2~eV) extracted from our extensive catalog. Iron atoms are depicted as blue spheres, while hydrogen atoms are represented by red spheres. Across all three topologies, the barrier linked to the displaced iron atom is denoted with an asterisk (*).}
    \label{figtopology}
\end{figure}

In the presence of H, we observe a number of barriers distributed between 1.6 and 1.9~eV that are associated with a broader spectrum of energies, indicating that the specific position of H atoms in the GB finely impacts the energy landscape. Most of these events are associated with the rearrangement of Fe atoms in the second and third-neighbor positions with respect to the GB. Fig.~\ref{figtopology} shows two  event topologies as examples from our comprehensive catalog, to offer deeper insights into the barriers that emerge when our GB is saturated with H atoms, which illustrates how the presence of saturated H leads to the diffusion of iron atoms over even greater distances from the GB. A detailed summary of these events is provided in Table~\ref{topology}. In the figure, the blue balls represent iron atoms and the red balls represent hydrogen atoms. The central diffusion Fe atom is marked with an asterisk (*), as shown in the previous figure. When hydrogen saturates the GB region, it becomes more difficult for Fe atoms to move within the GB, leading to an increased diffusion barrier, as mentioned previously. However, Fe atoms located in the second and third nearest-neighbor positions to the GB can still move towards it. These movements require crossing lower-energy barriers compared to Fe atoms within the GB, resulting in a higher rate of occurrence.

\begin{table}[!tb]
\caption{\label{topology}
Details of the events depicted in Fig.~\ref{figtopology}. Barrier, event rate, and Root-mean square displacement computed as in Table~\ref{tab:table2}.
}
\begin{ruledtabular}
\begin{tabular}{cccc}
Event &\scriptsize{ \begin{tabular}{@{}c@{}}Barrier \\ (eV)\end{tabular}}& $\Gamma_{if}$ (Hz) & $d_{si}$ 
(\AA)\\
\hline
\\
Event 1 & 1.63 & 4.19 $\times 10^{-15}$ & 1.944 \\
\\
Event 2 & 1.897 & 1.328 $\times 10^{-19}$ & 2.291 \\
\\

\end{tabular}
\end{ruledtabular}
\end{table}

\begin{figure*}[!htb]
    \centering
    \includegraphics[scale = 0.45]{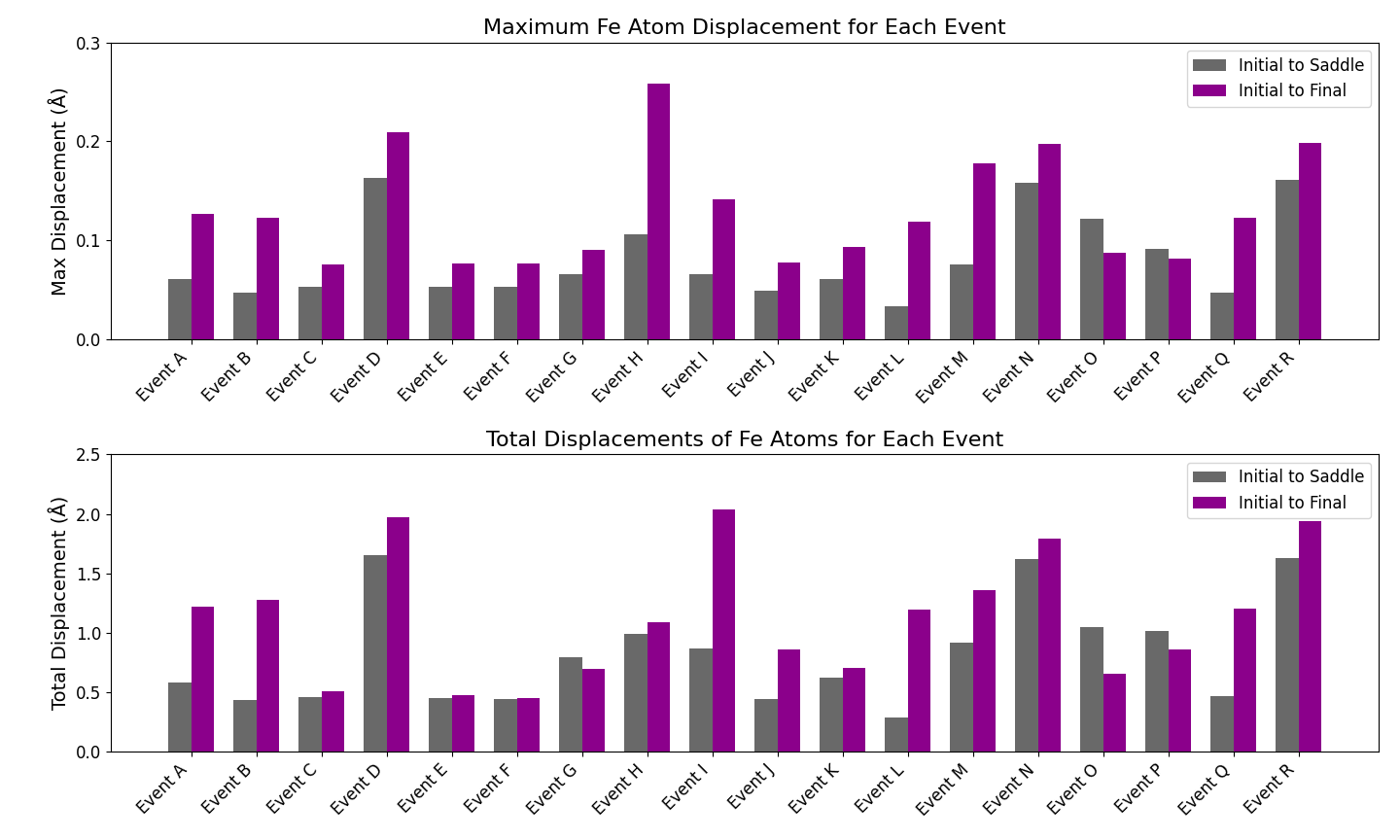}
    \caption{Maximum and total atomic displacement for H-dominated events in the H-saturated $\Sigma 37$ GB. These figures show the subset of H-dominated events with the highest displacement for an Fe. Top panel: maximum displacement (in \AA) for a Fe atom between the initial and saddle point (grey bars) and the initial and final point (purple bars). Bottom panel: total displacement of Fe atoms.}
    \label{displacementsigma37}
\end{figure*}

In addition to investigating the structural properties of the GB with and without H atoms, our study examines the behavior of H atoms within these boundaries. Our main objective is to determine whether the movement of Fe atoms during H-dominated events can lead to the diffusion of Fe atoms. For each event, which can involve up to 65 atoms, we calculate the displacement of the atom that moves the most during each event. We also calculate the displacement of all the Fe atoms from the initial to the saddle and from the initial to the final positions for these H-dominated events. The results are shown in Fig.~\ref{displacementsigma37}. In the upper panel, the displacement from the initial position to the saddle position is represented by gray, whereas the displacement from the initial position to the final position is indicated by violet. As shown, the H event with the largest Fe motion involves only a small Fe displacement, with the largest being approximately 0.25~\AA. The total displacement summed over all the involved Fe atoms is also shown at the bottom part of this figure. Although the largest total displacement is close to 1.97~\AA, this event involves a large number of Fe atoms that are shifted from their initial position without any barrier crossing. These findings indicate that, as H atoms move around the GB, they may involve considerable local relaxation of the surrounding Fe atoms.

\subsubsection{$\Sigma 3$ GB}

We now focus on the $\Sigma 3$ GB, specifically the symmetric $\Sigma 3(112)\theta=70.53^{\circ}$ tilt GB(Fig.~\ref{fig1}(b)). This particular GB configuration, characterized by a tilt angle of $70.53^{\circ}$, has been studied experimentally~\cite{lejcek2010grain}. Experimental studies have focused on the  $70.53^{\circ}$ $\langle 110 \rangle$ GB due to its unique structural properties, notably featuring a single structural unit that results in a sharp minimum in the GB energy~\cite{lejcek2010grain}. This distinctive characteristic makes the $\Sigma 3$ GB an excellent subject for further exploration of the influence of hydrogen on GBs.

As for the $\Sigma 37$ GB, we first characterize the GB's stability in its pure form without hydrogen to validate previous work, such as Ref.~\onlinecite{PhysRevB.97.054309}. Pure iron $70.53^{\circ}$ $\langle 110 \rangle$ GB demonstrates a higher stability than the $\Sigma 37$ GB and is characterized by $\sim$ 4~eV energy barriers for Fe atoms within the GB. Crossing these barriers leads to shallow metastable minima with inverse barriers of approximately 0.1~eV, favoring a return to the initial configuration of the ground state and improving the overall stability of the structure. 

\begin{figure}[!b]
    \centering
    \includegraphics[scale = 0.34]{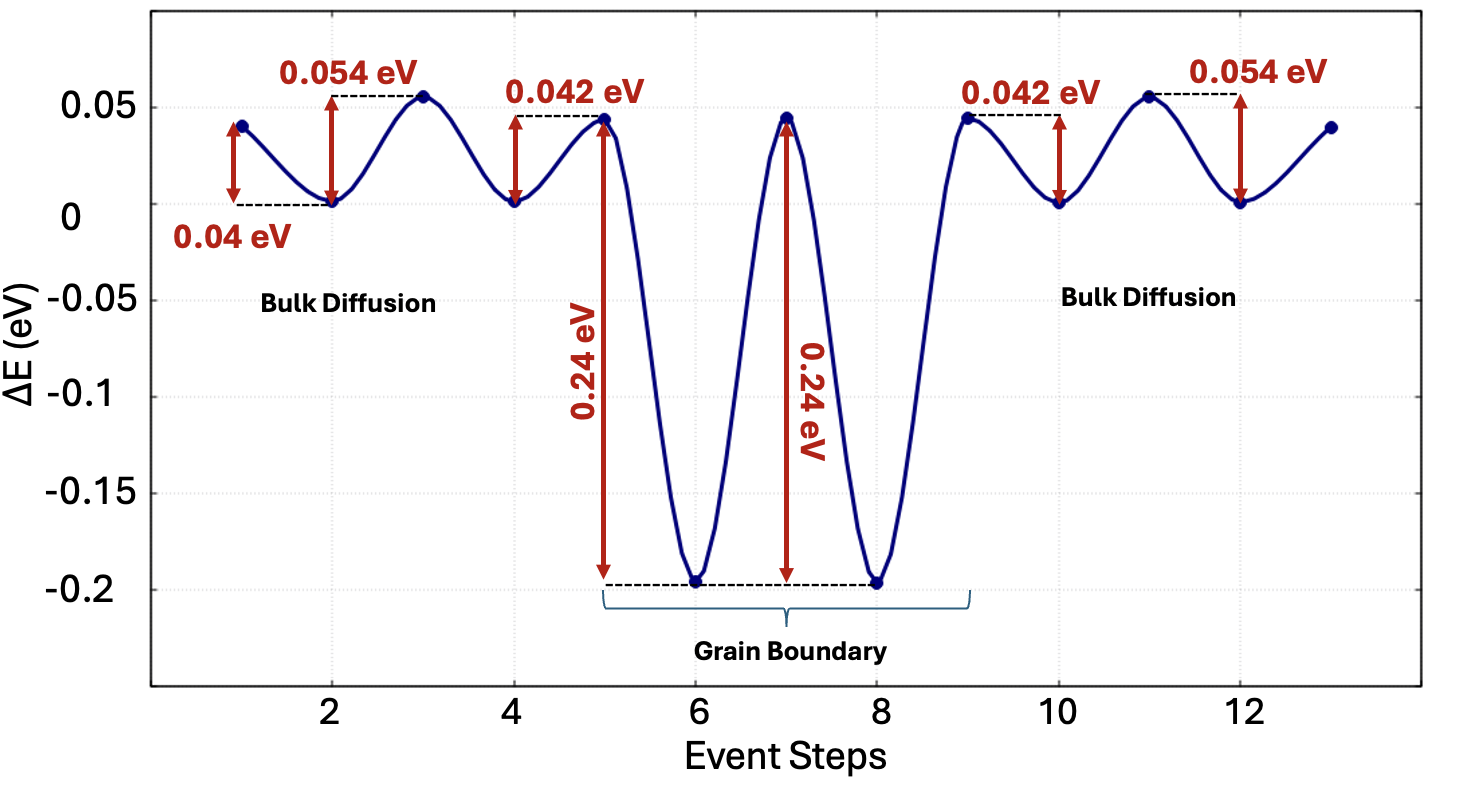}
    \caption{Energy landscape for a single H atom diffusing from the bulk to $\Sigma 3$ GB and within it. We observe only minimal elastic deformation around this type of GB. Motion of H within the GB or jumping out of it is achieved through a similar  0.24~eV energy, while bulk diffusion takes place by jumping over a 0.04~eV barrier from the second nearest neighbour.}
    \label{Trapping1H_sigma3}
\end{figure}

To turning our attention to H, we first examine the diffusion of a single H atom within this GB. H diffuses across barriers between 0.04~eV and 0.24~eV (Fig.~\ref{Trapping1H_sigma3}). Our study identifies 0.24~eV as the dominant barrier for H diffusion within $\Sigma 3$ GB. We note little elastic deformation near the GB, and the diffusion barriers vary little as H moves towards the GB, with a maximum barrier of 0.054~eV in the second-neighbor position. Trapping stabilizes the H by approximately 0.2~eV. However, from this position, a similar energy barrier (0.24~eV) must be overcome for H to move into the GB or through the bulk. This suggests that H-diffusion takes place through the bulk and that, while the energetics favor H being trapped in the GB, the absence of long-range deformations favors H moving rapidly into the GB. 

As for the first GB, we launch LAMMPS MD simulations to establish its H saturation levels. The saturation point occurs with 504 H atoms, representing 11.1~at\% atomic percent within the total structure. This represents a higher concentration of H atoms than for the $\Sigma 37$ GB. The interface energy of the GB, the solution energy of the H atoms in both the bulk and the GB, and the segregation energy are presented in Table~\ref{tab:table3}. In contrast to the $\Sigma 37$ GB, the segregation energy for the $\Sigma 3$ GB is positive for the saturated system, indicating a propensity for H at high concentrations to segregate away from the GB interface.

Similar to our study of the $\Sigma 37$ GB, starting from this saturated structure, we explore the diffusion mechanisms for Fe atoms surrounding the GB. For this geometry, each atom in the initial structure of the pure GB is associated with one of the five topologies, whereas we identify eight different topologies for the H-saturated system. As in previous work~\cite{PhysRevB.97.054309}, we find that without hydrogen atoms present, the diffusion barriers for Fe atoms range from 4.36 to 4.92~eV. Focusing on Fe atoms, with the same parameters, k-ART generates 9,070 events in the pure system and 1,512 Fe-dominated events in the H-saturated system. 

\begin{figure}[!b]
    \centering
    \hspace*{-0.3cm}\includegraphics[scale=0.29]{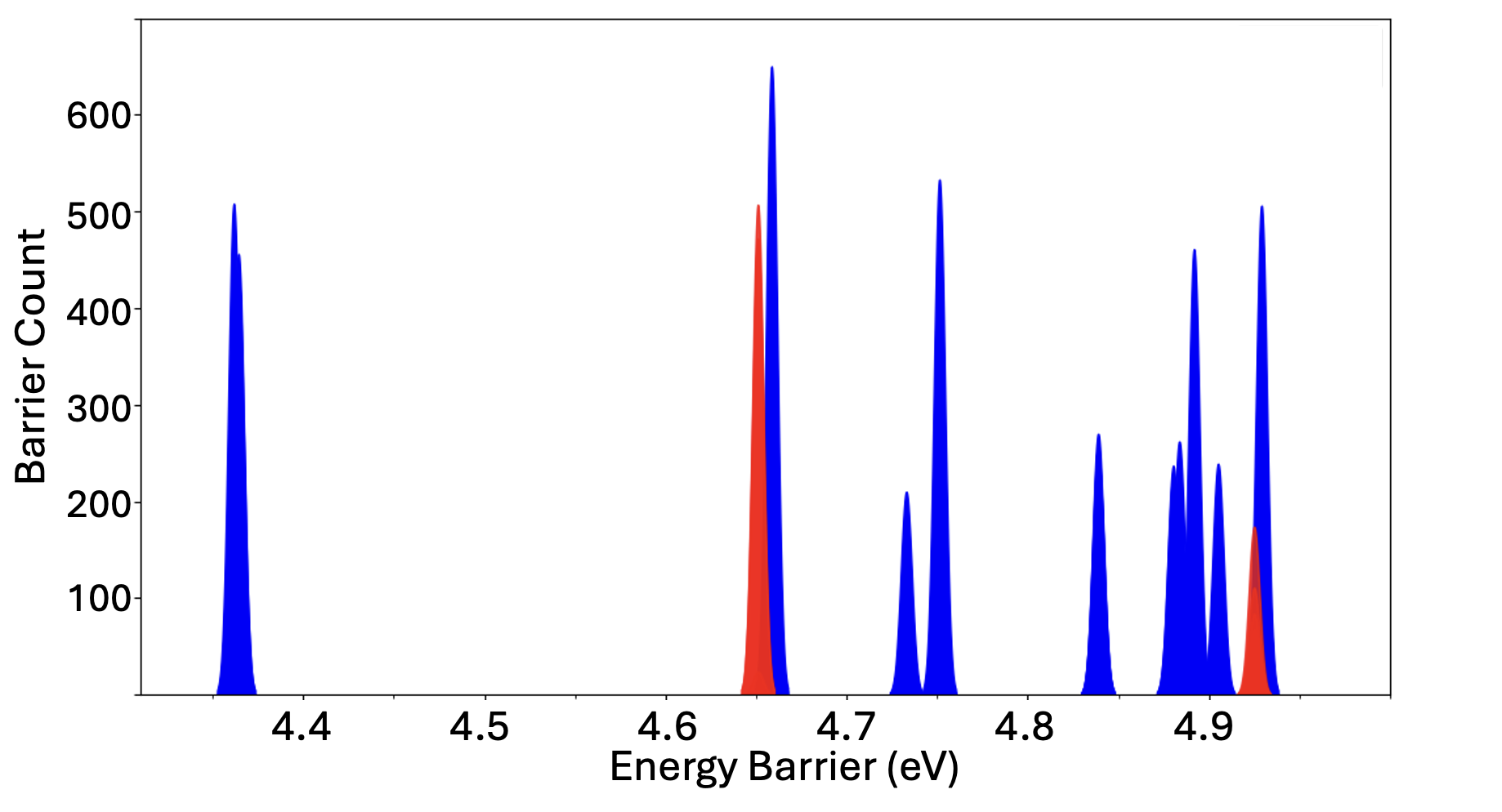}
    \caption{Energy-barrier distribution of the 9,070 events centered on Fe generated in the pure $\Sigma 3$ GB and of the 1,512 events also centered on Fe but generated in the H-saturated GB as a function of energy barrier. Diffusion barriers for Fe atoms in the pure system are represented in blue, whereas red represents barriers for the H-saturated GB.}
    \label{fig2.1}
\end{figure}

As for the $\Sigma 3$ GB, the presence of H brings important changes to the energy landscape (Fig.~\ref{fig2.1}): the lowest-energy barrier, at 4.36~eV disappears, along with a few others below 5~eV, with a clear rigidification of the system: the barriers are shifted to higher values, with the lowest mechanism now at 4.65~eV.

\begin{figure}[!tb]
    \centering
    \includegraphics[scale = 0.28]{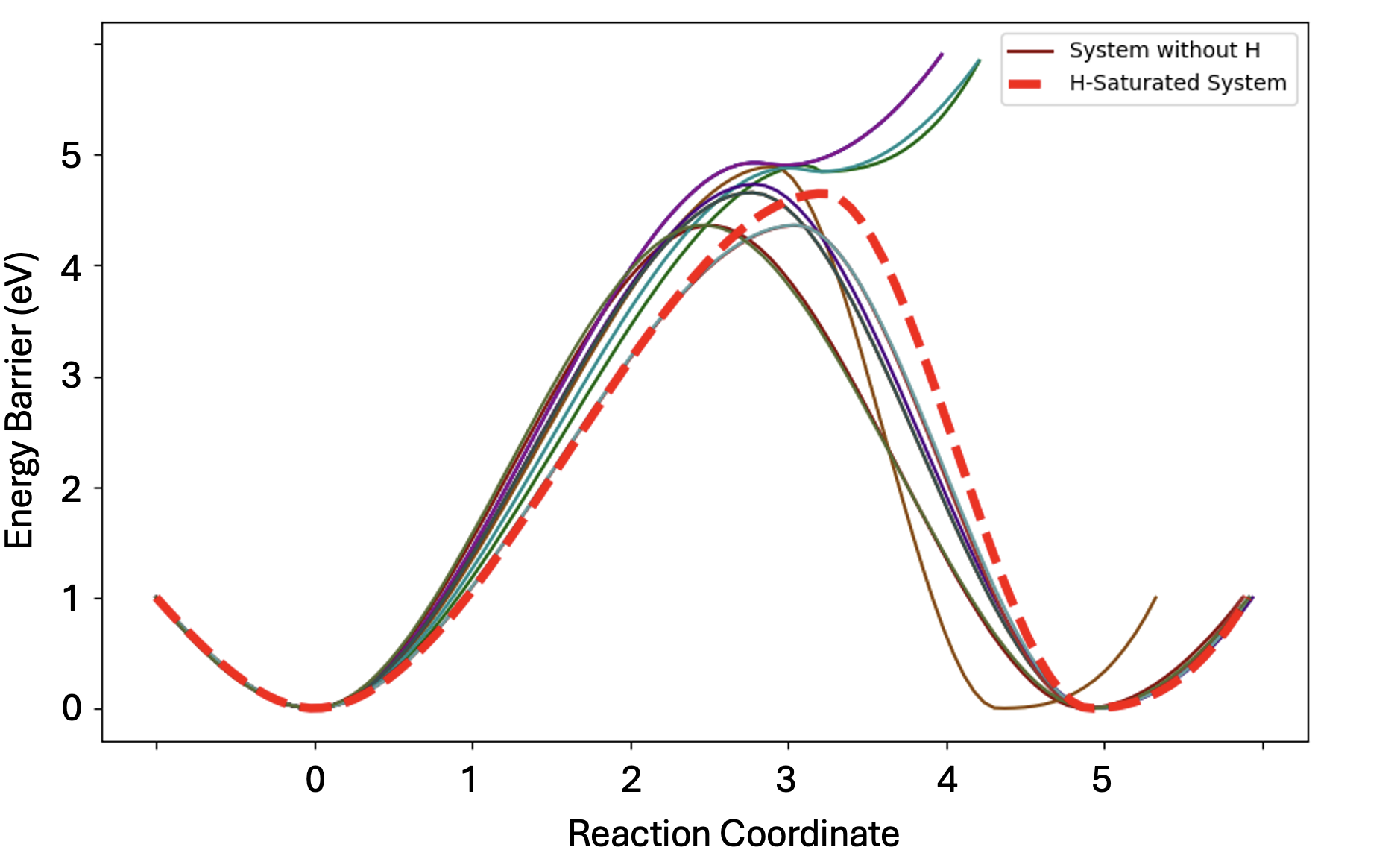}
    \caption{Energy pathways for the Fe diffusion mechanisms present in this figure.  The solid lines represent the possible energy barriers of Fe-event without H, while the dashed red line represents the energy barrier of Fe atom in the H-saturated system. In the pure GB, each event can have up to 11 possible pathways, with inverse barriers ranging from as high as 4.8~eV to as low as 0.26~eV. However, when H atoms are introduced, the lowest barrier for Fe atom diffusion, at 4.65~eV, becomes almost symmetric, leading to a slightly more stable state, 0.19~eV below. }
    \label{InverseBarrier}
\end{figure}

The nature of these events has also been modified. For pure GB, each event can have up to 11 possible pathways, with inverse barriers ranging from as large as 4.8~eV to as small as 0.26~eV. However, when H atoms were introduced into the system, the only possible barrier for Fe atom diffusion becomes symmetric. This is visible in Fig.~\ref{InverseBarrier}, which shows the energy pathways for the lowest energy events in both the pure and the H-saturated systems. 

\begin{figure*}[!tb]
    \centering
    \includegraphics[scale = 0.44]{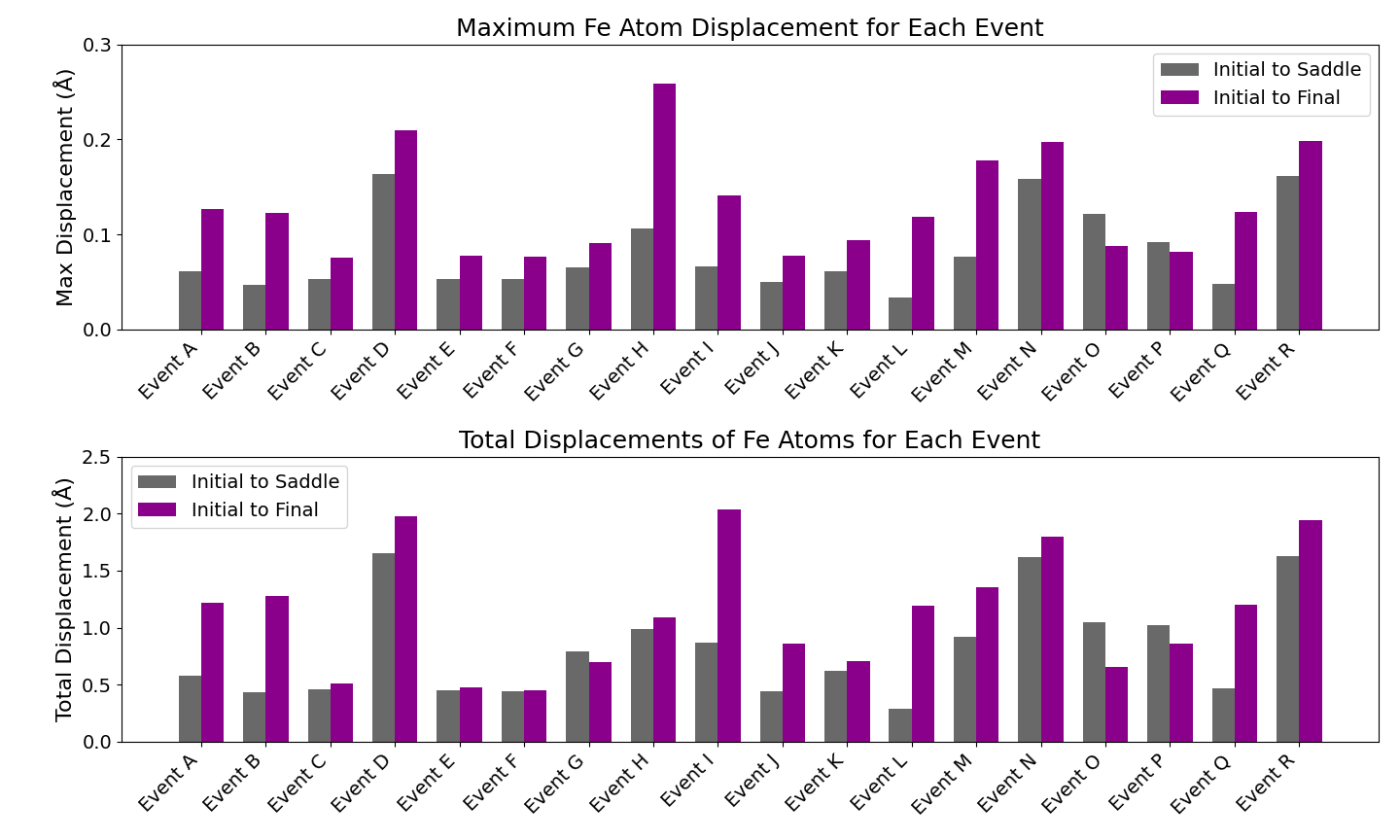}
    \caption{Maximum and total atomic displacement for H-dominated events in the H-saturated $\Sigma 3$ GB. These figures shows the subset of H-dominated events with the highest displacement for an Fe. Top panel: maximum displacement (in \AA) for a Fe atom between the initial and saddle point (grey bars) and the initial and final point (purple bars). Bottom panel: total displacement of Fe atoms.}
    \label{displacementsigma3}
\end{figure*}

As with the $\Sigma 37$, we study H-events to determine whether the movement of Fe atoms during these events leads to the diffusion of H atoms. Our results are shown in Fig.~\ref{displacementsigma3}. The Fe atom that moves the most displaces around 0.37 Å. The total displacement for all Fe atoms in each event is also shown, with the highest total displacement reaching approximately 2.55 Å. Similar to our findings for the previous GB, H-based events cause subtle movements and rearrangements of Fe atoms within the GB, without significant diffusion. Comparing these displacements with those observed in $\Sigma 37$, we notice slightly more deviations in the $\Sigma 3$ GB.

\section{\label{discussion}Discussion and Conclusion}
\label{section:Discussion}

In this study, we investigate the impact of hydrogen (H) on the kinetics of iron (Fe) atoms in two distinct grain boundaries (GB), namely $\Sigma 37$ and $\Sigma 3$. With a detailed exploration of the energy landscape of two GBs in the presence of H atoms, this study aims to help understand the HE mechanisms associated with the Hydrogen-Enhanced-Decohesion (HEDE) which was initially proposed by Pfeil~\textit{et al.} in 1926~\cite{doi:10.1098/rspa.1926.0103} and was later theoretically refined by Troiano~\cite{troiano1960role}, or Hydrogen-Enhanced-Localized-Plasticity (HELP)~\cite{BIRNBAUM1994191} mechanisms in metals. 

More specifically, we employ the kinetic activation relaxation technique (k-ART)~\cite{el2008kinetic,beland2011kinetic} with an EAM-based potential~\cite{song2013atomic} andgenerate extensive catalogs of activated events for all atoms in these structures in the absence and presence of H. Our previous results~\cite{khosravi2023kinetics} demonstrated that the empirical forcefield used here, which is essential for the large systems studied, yields results comparable to ab initio calculations, providing additional confidence in the qualitative and quantitative results presented here.

To assess the role of H, we first establish the stability of the two GBs in the absence of H. We then characterize the diffusion and behavior of one hydrogen atom in the presence of these GBs. We then turn to the study of the impact of fully H-saturated GBs, validating the critical concentration using molecular dynamics. 

The first GB studied is a $\Sigma 37$ system, a structurally stable configuration, with dynamics predominantly characterized by transitions between the ground state and less stable higher energy configurations, typically situated 1.4 to 1.6~eV above the ground state. Building on this groundwork, we then turn to the diffusion behavior of a single H atom within this GB. 
As we previously determined~\cite{khosravi2023kinetics}, the barrier for bulk diffusion to be 0.04~eV. $\Sigma37$, 
the trapping of the H, at 0.394~eV below the bulk, requires the atom to overcome successive barriers of 0.055~eV, 0.178~eV, and 0.077~eV (Fig.~\ref{Trapping1H_sigma37}). 

For its part, the $\Sigma 3$ exhibits higher stability in its pure form without hydrogen compared to the $\Sigma 37$GB, with energy barriers for Fe atoms within the GB characterized by approximately 4~eV. By studying the diffusion of the H at this GB, the $\Sigma 3$ GB system imposes very limited elastic deformations on the crystal. Therefore, the H-diffusion barriers near the GB are almost identical to those in the bulk. The highest barrier on the path to trapping, at 0.198~eV below the bulk, is only 0.054~eV as shown in Fig.~\ref{Trapping1H_sigma3}. 

The barrier for a H atom to detach from a vacancy is nearly the same (0.54~eV) as the barrier for detaching from the $\Sigma 37$ GB (0.52~eV)~\cite{khosravi2023kinetics}. However, once in the GB, H diffuses along the GB by crossing an effective barrier of 0.22~eV, which is a low-barrier at room temperature, but is still much higher than that in the bulk. In the case of the $\Sigma 3$ GB, the barrier for H to detrap from the GB is 0.24~eV, significantly lower than the barrier observed for H trapping in the vacancy. Moreover, the detrapping barrier has the same height as that in GB diffusion. The restraining effect observed in the interaction between vacancies and H is therefore similar for the $\Sigma 37$ case, where H tends to remain trapped, whereas the second GB ($\Sigma 3$) is less constrained, allowing hydrogen to detach and diffuse more readily into the bulk.

After establishing the optimal hydrogen saturation level in the GB, we turn to assessing the effect of this high concentration of H on the kinetics of the GB by comparing the activated mechanisms for Fe atoms in pure and H-saturated GBs. 

The two GBs show very different energy landscapes. Limiting the analysis to barriers below 5~eV, the pure $\Sigma 37$ GB shows a rich landscape, with hundreds of different mechanisms distributed almost continuously between 1.48 and 5~eV, while the pure $\Sigma 3$ GB, which is much more stable, displays limited activated mechanisms, with barriers between 4.36 and 4.92~eV.

However, the effect of H saturation on these energy landscapes is similar for both the GBs (Figs.~\ref{sigma37_distribution} and ~\ref{fig2.1}). First, by rigidifying the local environment, H-saturation shifts the activation barriers for Fe-dominated events to higher energies, thereby stabilizing the GB configuration. For the $\Sigma 37$ GB, the lowest-energy Fe-dominated real diffusion mechanism found in the pure system at 1.48~eV ~\cite{PhysRevB.97.054309} is pushed to 1.74~eV. Similarly, for the $\Sigma 3$ GB, the lowest barrier shifted from 4.36 to 4.65~eV.

Beyond energy shifts, saturating the GBs greatly reduces the number of activation mechanisms available: from 21,482 to 8,368 events in the case of the $\Sigma 37$ GB, 62~\% drop; and from 9,070 to 1,512 events, a 82~\% reduction for the $\Sigma 3$ GB. For the second GB, this drop is associated with the disappearance of the mechanisms, leading to very metastable states. This effective smoothing of the energy landscape, therefore, leaves more symmetric and predominantly diffusive mechanisms. 

In the case of the $\Sigma 37$ GB, a more detailed analysis also shows that the presence of H, therefore, causes significant elastic deformation around the GB and affects the diffusion of Fe atoms not only directly at the GB but also in the second neighbor position. Thus, H atoms introduce new diffusion pathways and increase the frequency of the diffusion barriers within this range.

As shown in Table~\ref{tab:table3}, the negative segregation energy of the $\Sigma 37$ GB indicates that H will preferably diffuse into the boundary. However, as H moves in, the boundary will stabilize and become slightly more rigid but still able to diffuse, albeit at a slightly lower rate. A positive segregation energy for H in the $\Sigma 3$ GB, means that H is less likely to settle in the GB. However, a H-saturated GB will have an effect similar to that  $\Sigma 37$ GB.

Although a clear experimental validation of the effect of hydrogen on atomic cohesion remains difficult to obtain, numerical studies have concluded that the presence of H reduces the cohesion energy within diverse GBs, with large differences in impact. For example, Momida \textit{et al.}~\cite{PhysRevB.88.144107}, reported a 4\% decrease in the $\Sigma 3(112)$ ideal strength, a result supported by Tahir \textit{et al.}~\cite{TAHIR2014462} who observed a 6\% reduction in strength with a monolayer of H at a coverage of one H atom per structural unit of the $\Sigma 5$ GB. On the other hand, Katzarov and Paxton~\cite{PhysRevMaterials.1.033603} calculated a much more substantial decrease from 33 GPa to 22 GPa with increasing H concentration, similar to that obtained by Wang \textit{et al.}'s~\cite{WANG2016279} comprehensive modeling, utilizing an empirical EAM potential across a spectrum of GBs, which predicted a significant 37\% drop in cohesive energy under conditions leading to intergranular fracture, thus supporting a contribution to decohesion. While this drop in cohesive energy is consistent with the HEDE, other mechanisms can be at play that involve plasticity, such as the postulated HELP mechanism.  

Here, at  comprehensive exploration of all potential pathways with k-ART generates detailed information that complements previous simulations and helps enrich these two proposed mechanisms. First, in both types of GBs, which were investigated in this study, hydrogen encounters higher diffusion barriers compared to the bulk material, indicating that hydrogen has lower diffusivity within these GBs than in the bulk. While many studies have postulated that GBs are conduits for fast hydrogen diffusion~\cite{LEE1986301,dieudonne2012sims, LADNA19872537}, our work supports other recent analyses that show a strong dependence on the specific local geometry of the GBs~\cite{PhysRevLett.122.215501}.

More importantly, for the purpose of embrittlement, the microscopic understanding gained here is about the evolution of Fe diffusion barriers at the grain boundary. Both the HEDE and HELP focused on the net macroscopic effect of H at the GBs. Here, we provide microscopic details regarding the effect of H on the energy landscape of the Fe atoms around the GBs. We find that H-saturated GB are rigidified through two mechanisms: first, the activation barriers for Fe diffusion are shifted to higher values (around 0.25eV), making motion slightly more difficult; second, the number of potential mechanisms is significantly reduced, by 62 and 82~\%, respectively, for the $\Sigma 37$ and the $\Sigma 3$ GBs, respectively. If the barrier shift is not sufficient to fundamentally change the cohesiveness of the interface, many of the remaining events involve the diffusion of Fe atoms in the second and third nearest neighbor positions from the GB, reducing the effective diffusivity of the GB interface. These observations show two competing and opposite trends: (i) H can increase the number of active atomic layers around the GB, potentially contributing to an increase in plasticity; and (ii)  at the same time, H contributes to rigidifying the interface by slightly increasing the barrier height for relaxation mechanisms and decreasing the number of those mechanisms. These observations indicate that both HEDE and HELP are intertwined at the microscopic level and that the balance between plasticity and decohesion energy is likely dependent on the specific GB geometry.


Overall, this study provides the first detailed examination of the changes in the energy landscape of GBs in BCC Fe in the presence of H. Initial results show that this characterization is essential for understanding hydrogen embrittlement and identifying a microscopic basis for dominant embrittlement mechanisms, such as HEDE and HELP. Our results form the basis for further work to better link the evolution at the microscopic level with the overall mechanical properties of Fe in the presence of hydrogen.

\section{Code availability}
The k-ART package is accessible upon request. To obtain access to the repository, please reach out to the authors directly.

\section{Acknowledgments}

This research received partial support from a grant provided by the Natural Sciences and Engineering Research Council of Canada (NSERC). The authors extend their gratitude to Calcul Québec and the Digital Research Alliance of Canada for their generous provision of the computational resources.
This research benefited from the robust OVITO software developed by A Stukowski~\cite{stukowski2009visualization}, which greatly aided in the analysis and visualization of atomic configurations. OVITO is available for access at http://ovito.org/.

\bibliography{apssamp}
\bibliographystyle{apsrev4-1}

\end{document}